\documentclass[twocolumn,epjc3]{svjour3}
\smartqed  
\usepackage{amsmath}
\usepackage{dsfont}
\usepackage{amssymb}
\usepackage{enumerate}
\usepackage{esint}
\usepackage{color}
\usepackage{mathtools} 
\usepackage[hyperfootnotes=true]{hyperref}
\usepackage[left=2.5cm,right=2.5cm,top=2.5cm,bottom=2.5cm]{geometry}
\usepackage{cite}
\usepackage{ulem}

\newcommand{\n}{\nonumber}

\renewcommand{\d}{\mathrm{d}}
\renewcommand{\i}{\rm i}

 \makeatletter
 \newcommand{\pback}[1]{{
   \let\@rrow=\leftarrowfill
   \mathchoice{\AIN@stemPullBack{#1}{\@rrow}}{\AIN@stemPullBack{#1}{\@rrow}}
     {\AIN@indxPullBack{#1}{\@rrow}}{\AIN@indxPullBack{#1}{\@rrow}}}
   \vphantom{#1}}
 \newcommand{\AIN@stemPullBack}[2]{
   \vtop{\mathsurround=0pt
   \ialign{##\crcr$\textstyle{#1}\strut$\crcr
     \noalign{\kern-0.4ex\nointerlineskip}{\tiny#2}\crcr}}}

 \newcommand{\AIN@indxPullBack}[2]{
   \vtop{\mathsurround=0pt
   \ialign{##\crcr\hfil$\scriptstyle{#1}$\hfil\crcr
     \noalign{\kern+0.4ex\nointerlineskip}{\tiny#2}\crcr}}}

\makeatother

\title{Mechanics of Isolated Horizons in Scalar-Tensor Theories}
\author{Shupeng Song\thanksref{e1,addr1,addr2} \and Yongge Ma\thanksref{e2,addr1}}
\thankstext{e1}{e-mail: songsp@mail.bnu.edu.cn}
\thankstext{e2}{e-mail: mayg@bnu.edu.cn}
\institute{Department of Physics, Beijing Normal University, Beijing 100875, China\label{addr1}
\and
Institute for Gravitation and the Cosmos \& Physics Department, Penn State, University Park, PA 16802, USA\label{addr2}}
\date{Received: date / Revised version: date}

\journalname{Eur. Phys. J. C}
\begin{document}
\maketitle
\begin{abstract}
  Based on the first-order action for scalar-tensor theories with the Immirzi parameter, the symplectic form for the spacetimes admitting a weakly isolated horizon as internal boundary is derived by the covariant phase space approach. The first law of thermodynamics for the weakly isolated horizons with rotational symmetry is obtained. It turns out that the Immirzi parameter appears in the expression of the  angular momentum of isolated horizon, and the scalar field contributes to the horizon entropy.
\end{abstract}

\section{Introduction}
Scalar-tensor theories (STT) are inspired by Mach's principle~\cite{bergmann1968comments,PhysRevD.1.3209,PhysRev.124.925} and belong to a class of modified gravity theories. In some cosmological models, the non-minimally coupled scalar field in STT may cause the acceleration of the universe and hence explain the issue of dark energy\cite{sen2001late,PhysRevLett.85.2236,PhysRevD.83.043521,PhysRevD.63.043504,PhysRevD.71.061501}. The nonperturbative quantization of STT~\cite{PhysRevLett.106.171301,PhysRevD.84.064040,Zhang:2011gn,PhysRevD.84.104045,Ma_2012} and their cosmological models~\cite{Zhang:2012ta,Zhang:2012em,Artymowski:2013qua} has also been carried out recently by extending the method of loop quantum gravity~\cite{Ashtekar:2004eh,rovelli2005quantum,thiemann2007modern,han2007fundamental}.
The thermodynamics of black hole (BH) is an important issue in physics. While the event horizon of a BH is a global notion not suitable for local physics~\cite{hawking1973large}, the notion of isolated horizon is quasi-locally defined~\cite{Ashtekar:1998sp} and widely applied to calculate observables in numerical simulation of near horizon geometry~\cite{Ashtekar:2001jb,Dreyer:2002mx}. Practically, an isolated horizon satisfies the minimal requirements to derive the zeroth law and the first law of BH thermodynamics in general relativity (GR)~\cite{Ashtekar:2000hw,Ashtekar:1999yj,Ashtekar:2001is}.

The aim of this paper is to study the Hamiltonian structure and thermodynamics of isolated horizon in STT. To match the loop quantum STT, we start with the first-order action of STT with the Immirzi parameter proposed in \cite{Zhou:2012ie}. The connection dynamical formalism derived from this action via Hamiltonian analysis is completely consistent with that derived from the geometrical dynamics by canonical transformations~\cite{PhysRevD.84.104045}. It should be noted that the thermodynamics of isolated horizons  with a non-minimally coupled scalar field was first studied in Ref.~\cite{Ashtekar:2003jh}. That theory could be regarded an a special case of the general STT that we are considering. By our general treatment, there exist new elements in both the symplectic structure and first law of isolated horizon in STT. Throughout the paper, we use the capital Latin letters $I,J,K,\cdots$ to denote the internal Lorentzian indices, and the spacetime indices are denoted by $a,\ b,\ c,\ \cdots$. 

\section{Symplectic form}
We consider the weakly isolated horizon (WIH) in STT. In comparison with the definition of WIH in GR, its definition in STT contains also the extra requirement on the non-minimally coupled scalar field $\phi$~\cite{Ashtekar:2003jh}, such that the horizon is in equilibrium. A 3-dimensional null hypersurface $\Delta$ equipped with an equivalence class $[l]$ of null normals $l^a$ of a spacetime with metric $g_{ab}$ and scalar field $\phi$ in STT is said to be a WIH if the following conditions hold.
\begin{enumerate}[(i)]
\item The topology of $\Delta$ is $S^2\times\mathds{R}$;
\item The expansion $\theta_{(l)}$ of $l$ vanishes on $\Delta$ for any null normal $l$;
\item The scalar field satisfies $\mathcal{L}_l\phi\,\widehat{=}\,0$, where $\mathcal{L}_l$ denotes the Lie derivative along $l$ and ``$\widehat{=}$" means ``equal on $\Delta$";
\item Equations of motion hold on $\Delta$;
\item The equivalence class $[l]$ of the future-directed $l$ is chosen by $l\sim l'$ if and only if $l'^a=C\,l^a$ for a positive constant $C$, such that there is a connection 1-form $w$ on $\Delta$ defined by $\nabla_{\pback{a}}l^b\ \widehat{=}\ w_a l^b$ and satisfying $\mathcal{L}_l w_a\ \widehat{=}\ 0$ for all $l\in[l]$, where $\nabla_{\pback{a}}$ denotes the pullback of the spacetime connection $\nabla_a$ compatible with $g_{ab}$ to $\Delta$.
\end{enumerate}
Hereafter, the pullback of a covariant index to $\Delta$ will be denoted by an arrow under that index.
By the definition, one can obtain the following useful properties of WIHs in STT.
\begin{enumerate}[(a)]
  \item The surface gravity $\kappa_{(l)}\ \widehat{=}\ w_a l^a$ is a constant on $\Delta$, which gives the zeroth law of its thermodynamics;
  \item There is a natural area 2-form $\prescript{2}{}\epsilon$ on $\Delta$ satisfying $\mathcal{L}_l\prescript{2}{}\epsilon\ \widehat{=}\ 0$ and $\prescript{2}{}\epsilon_{ab}l^b\ \widehat{=}\ 0$;
  \item One can define the potential $\psi$ of the surface gravity by $\mathcal{L}_l\psi\ \widehat{=}\ \kappa_{(l)}$.
  \item There is a unique induced covariant derivative $\mathcal{D}$ inherited from $\nabla$. The actions of $\mathcal{D}$ on a vector field $X^a$ tangent to $\Delta$ and on an 1-form $Y_a$ intrinsic to $\Delta$ are given
by $\mathcal{D}_aX^b\widehat{=}\nabla_{\pback{a}}\tilde{X^b}$ and $\mathcal{D}_aY_b\widehat{=}\pback{\nabla_{a}\tilde{Y}_b}$ respectively, where $\tilde{X}^b$ and $\tilde{Y}_b$ are the arbitrary extensions of $X^a$ and $Y_a$ to the 4-dimensional spacetime.
\end{enumerate}

Now we follow the covariant phase space approach \cite{Lee:1990nz,Ashtekar:1991} to derive the symplectic structure of WIH in STT.
The first-order action of STT in \cite{Zhou:2012ie} can be written in terms of differential forms as
\begin{align}
  S[e,\bar{\omega},\phi]=&\frac{1}{16\pi G}\int_{M} \phi \Sigma^{IJ}\wedge\bar{\Omega}_{IJ}-\Sigma^{IJ}\wedge\bar{\omega}_{IJ}\wedge \d\phi\n\\
  &+\Sigma^{IJ}\wedge \d V_{IJ}+\frac{\omega(\phi)}{\phi}\prescript{\ast}{}\! \d\phi\wedge \d\phi\n\\
  &+\frac{1}{\gamma}\Sigma^{IJ}\wedge\prescript{\star}{}{\bar{\Omega}}_{IJ}+\xi(\phi)\epsilon, \label{action}
\end{align}
where $G$ is the gravitational constant, $\Sigma^{IJ}_{ab}:=\frac{1}{2}{\varepsilon^{I\!J}}_{\!\!K\!L}\newline e^K_a\wedge e^L_b$ with $e^I_a$ and $\varepsilon_{IJKL}$ being the cotetrad and Levi-Civita symbol respectively, $(\d V_{IJ})_{ab}=\partial_{\left[a\right.}e^I_{\left.b\right]} e^{cJ}\partial_c\phi$, the curvature 2-form of the Lorentzian connection $\bar{\omega}^{IJ}_{a}$ is defined by $\bar{\Omega}^{IJ}_{ab}=\d_a\bar{\omega}^{IJ}_b+\bar{\omega}^{\;I}_{a\;K}\wedge\bar{\omega}_b^{KJ}$ , $\prescript{\star}{}{\bar{\Omega}}_{IJ}=\frac{1}{2}\varepsilon_{IJKL}\bar{\Omega}^{KL}$, $(\prescript{\ast}{} \d\phi)_{abc}=(\d_e\phi){\epsilon^e}_{abc}$ with $\epsilon$ being the spacetime volume element, the potential $\xi(\phi)$ and the coupling parameter $\omega(\phi)$ are general functions of $\phi$, and the Immirzi parameter $\gamma$ is included. It has been shown in Ref.~\cite{Zhou:2012ie} that action \eqref{action} gives the same field equations with the general second-order action of scalar-tensor theories. Our motivation to choose action \eqref{action} is that the connection dynamics can be derived from it, which is the foundation of loop quantum scalar-tensor theories~\cite{PhysRevD.84.104045}. It should be noted that the general scalar-tensor theories that we are considering do not include those extended theories containing the higher-order derivatives of $\phi$ in their actions, such as Horndeski theory~\cite{Horndeski:1974wa} and GLPV theory~\cite{Gleyzes:2014dya}.

Let us consider a spacetime region $\mathcal{M}$ which admits a WIH $(\Delta,[l])$ as its internal boundary and is bounded by two spatially partial Cauchy surfaces $M^{\pm}$ intersecting $\Delta$ at two-spheres $H^{\pm}$ and extending to spatial infinity $i^0$. The asymptotic conditions on $i^0$ are that the metric is asymptotic flat and $\phi\to 1$. The boundary of spacetime region $\mathcal{M}$ is $\partial\mathcal{M}=\Delta\cup M^+\cup M^-\cup i^0$.  To satisfy the variational principle, one has to add a boundary term at $i^0$ such that action \eqref{action} becomes
\begin{align}\label{action-2}
  S[e,\bar{\omega},\phi]\!=\!&\frac{1}{16\pi G}\int_{M} \phi \Sigma^{IJ}\wedge\bar{\Omega}_{IJ}-\Sigma^{IJ}\wedge\bar{\omega}_{IJ}\wedge \d\phi\n\\
  &+\!\Sigma^{IJ}\wedge \d V_{IJ}+\frac{\omega(\phi)}{\phi}\prescript{\ast}{}\! \d\phi\wedge \d\phi\n\\
  &+\!\frac{1}{\gamma}\Sigma^{IJ}\wedge\prescript{\star}{}{\bar{\Omega}}_{IJ}+\xi(\phi)\epsilon\n\\
  &-\!\frac{1}{16\pi G}\int_{i^0}\!\!\Sigma^{IJ}\!\wedge\!\left(\phi\,\bar{\omega}_{IJ} \!+\!\frac{1}{\gamma}\prescript{\star}{}{\bar{\omega}}_{IJ}\!\!\!+\!e_I e^f_J\partial_f\phi\right)
\end{align}
The variation of action \eqref{action-2} reads
\begin{align}
  \delta S[\Psi]=\int_{\mathcal{M}} E[\Psi]\;\delta\Psi+\int_{\partial \mathcal{M}-i^0}\mathcal{J}[\Psi,\delta\Psi], \nonumber
\end{align}
where $\Psi$ denotes the fields $e$, $\bar{\omega}$ and $\phi$, $E[\Psi]=0$ is the equation of motion for $\Psi$, and the current 3-form $J$ reads
\begin{align}
  &\mathcal{J}[\Psi,\delta\Psi]=\frac{1}{16\pi G}\left[\Sigma^{IJ}\!\wedge\delta\left(\phi\,\bar{\omega}_{IJ} \!+\!\frac{1}{\gamma}\prescript{\star}{}{\bar{\omega}}_{IJ}\!+\!e_I e^f_J\partial_f\phi\right)\right.\n\\
  &\left.\left(\!\frac{\omega(\phi)}{\phi}\prescript{\ast}{} \d\phi\!-\!\frac{1}{3}\varepsilon_{I\!J\!K\!L}(\nabla_d e^{dI}\!)e^J\!\!\!\wedge\! e^K\!\!\!\wedge\! e^L\!\!\!\right)\delta\phi\right]. \label{boundaryterm}
\end{align}
Note that the potential $\xi(\phi)$ does not contribute to $J$. Comparing with the current 3-form in \cite{Ashtekar:2003jh}, the boundary term \eqref{boundaryterm} contains also the scalar field part. We will see that the scalar part in the second term in the right hand side of Eq.~\eqref{boundaryterm} also contributes to the boundary symplectic structure. Note that, different from that in \cite{Ashtekar:2003jh}, the connection $\bar{\omega}$ is compatible with the tetrad $e$ in our covariant phase space $\Gamma$.


The (pre-)symplectic form can be obtained by the second variation of the action~\cite{Ashtekar:2000hw}. First, the anti-symmetrized second variation gives the symplectic current $J$ on a point $p$ in the covariant phase space $\Gamma$. $J$ is a closed 3-form on $\mathcal{M}$, since the fields $\Psi$ satisfy the equations of motion and  the variations $\delta\Psi$ satisfy the linearized equations off $p$. Second, the asymptotic conditions guarantee that the integral of symplectic current over $i^0$ vanishes. Therefore the integral of $J$ over a partial Cauchy surface $M$ and its intersection with $\Delta$ is conserved, which can be defined as a pre-symplectic form.
The symplectic current $J$ in our situation satisfies
\begin{align}\label{current1}
  &8\pi G\,J(p,\delta_1,\delta_2)\n\\
  =&\delta_{\left[1\right.}\Sigma^{IJ}\wedge\delta_{\left.2\right]}\left(\phi\,\bar{\omega}_{IJ} +\frac{1}{\gamma}\prescript{\star}{}{\bar{\omega}}_{IJ}+e_I e^f_J\partial_f\phi\right)\n\\ -&\delta_{\left[1\right.}\left(\frac{\omega(\phi)}{\phi}\prescript{\ast}{} \d\phi\!-\!\frac{1}{3}\varepsilon_{IJKL}(\nabla_d e^{dI})e^J\!\!\!\wedge\! e^K\!\!\!\wedge\! e^L\!\!\!\right)\delta_{\left.2\right]}\phi.
\end{align}
For the spacetime region $\mathcal{M}$, the closed condition of $J$ leads to
\begin{align}
  \left(\int_{M^+}-\int_{M^-}+\int_{\Delta}\right) J(p,\delta_1,\delta_2)=0. \label{intJ}
\end{align}
Note that $\mathcal{M}$ is topologically $M\times\mathds{R}$, where $M$ is an oriented spatial manifold with an internal 2-sphere boundary $H$. It is convenient to fix an internal null tetrad $(l^I,n^I,m^I,\bar{m}^I)$ on $\Delta$ such that each element of the tetrad is annihilated by a fiducial flat connection $\partial$. The conditions of WIH ensure that one could choose the tetrads $e^a_I$ such that their contraction with the internal null tetrad could give the spacetime null tetrad $(l^a,n^a,m^a,\bar{m}^a)$ adapted to $\Delta$ and $M$ \cite{Ashtekar:2000hw}. This means that $l^a$ belongs to $[l]$ fixed on $\Delta$, the complex null vector $m^a$ is tangential to $H$, $n^a$ is future directed and transverse to $\Delta$, and they satisfy $n^a l^b g_{ab}=-1$, $m^a \bar{m}^b g_{ab}=1$ and all other scalar products vanishing. To calculate the symplectic current on $\Delta$, it is useful to express $J$ by the null tetrads. The pull-back of the two-forms $\Sigma^{IJ}$ to $\Delta$ can be expressed as \cite{Ashtekar:2000hw}

\begin{align}\label{sigma}
  \underleftarrow{\Sigma}^{IJ}\ \widehat{=}\ 2l^{\left[I\right.}n^{\left.J\right]}\prescript{2}{}\epsilon +2n\wedge(iml^{\left[I\right.}\bar{m}^{\left.J\right]}-i\bar{m}l^{\left[I\right.}m^{\left.J\right]}),
\end{align}
where $\prescript{2}{}\epsilon=im\wedge\bar{m}$ is the area 2-form on the 2-sphere $H$. The pull-back of the connection $\bar{\omega}_a^{IJ}$to $\Delta$ can be expressed by \cite{Chatterjee:2008if}
\begin{align}\label{connection}
  \pback{\bar{\omega}}^{IJ}=&-2l^{\left[I\right.}n^{\left.J\right]}w +2m^{\left[I\right.}\bar{m}^{\left.J\right]}V +2l^{\left[I\right.}\bar{m}^{\left.J\right]}U \n\\ &+2l^{\left[I\right.}m^{\left.J\right]}\bar{U},
\end{align}
where the components of 1-forms $\omega$, $V$ and $U$ can be expressed by the Newman-Penrose spin coefficients as
\begin{subequations}\label{com}
\begin{align}
w&=-\kappa_{(l)} n+(\alpha+\bar{\beta})m+(\bar\alpha+\beta)\bar{m}, \label{com-ome}\\
  V&=-(\epsilon-\bar{\epsilon})n+(\alpha-\bar{\beta})m+(\beta-\bar{\alpha})\bar{m},\label{com-V} \\
U&=-\bar{\pi}n+\bar{\tilde{\mu}}m+\bar{\lambda}\bar{m}.
\end{align}
\end{subequations}
Now we consider the terms containing the scalar field in Eq.~\eqref{current1}.
Because the pull-back of the covariant index annihilates the $l_a$ direction and the scalar field satisfies the condition (iii) of WIH, the pull-back of the one-form $e_{cI}e_J^f\partial_f\phi$ can be written as
\begin{align}\label{extra1}
  e_{\pback{c}I}e_J^f\partial_f\phi=&(-l_I n_c+\bar{m}_I m_c +m_I\bar{m}_c)\times\n\\
  &(-l_J n^f+\bar{m}_J m^f+m_J\bar{m}^f)\partial_f\phi.
\end{align}
Since the right hand side of Eq.~\eqref{extra1} does not contain $n_I$ or $n_J$ components, its contraction with Eq.~\eqref{sigma} is zero. The pull-back of $\prescript{\ast}{} \d\phi$ to $\Delta$ reads
\begin{align}\label{dphi}
  (\prescript{\ast}{} \d\phi)_{\pback{abc}} =&(\d_d\phi){\epsilon^d}_{\pback{abc}}\n\\
  =&-\varepsilon_{IJKL}e_{\pback{a}}^I\!\wedge e_{\pback{b}}^J\!\wedge e_{\pback{c}}^K e^{dL}\d_d\phi.
\end{align}
Since the tetrads appeared in the right hand side of Eq.~\eqref{dphi} do not contain the $n^I$ components, their contraction with $\varepsilon_{IJKL}$ is zero.
Taking account of the identities $\varepsilon_{IJKL}l^I n^J m^K\bar{m}^L=\i$ and $\nabla_a l^a=\kappa_{(l)}$, the pull-back of the three-form $\frac{1}{3} \varepsilon_{IJKL} (\nabla_d e^{dI}) \cdot e^J\wedge e^K\wedge e^L$ reads
\begin{align}\label{extra2}
  &\frac{1}{3}\varepsilon_{IJKL}(\nabla_d e^{dI})e_{\pback{a}}^J\!\wedge e_{\pback{b}}^K\!\wedge e_{\pback{c}}^L\n\\
  &\qquad=2(\nabla_d l^d)\epsilon_{IJKL}n^I l^J m^K\bar{m}^L n_a\wedge\bar{m}_b\wedge m_c\n\\
  &\qquad=2\kappa_{(l)}n_a\wedge i\,m_b\wedge\bar{m}_c\n\\
  &\qquad=2\kappa_{(l)} \prescript{3}{}\epsilon,
\end{align}
where $\prescript{3}{}\epsilon$ denotes the three-volume element on $\Delta$.
It should be noted that the variations in Eq.~\eqref{current1} do not change the WIH $(\Delta,[l])$. Therefore the restriction to $\Delta$ and the variation of $\Psi$ commutate with each other. For instance, one has
\begin{align}
  \delta(\Sigma_{\pback{ab}}^{IJ})={(\delta\Sigma)}_{\pback{ab}}^{IJ}. \n
\end{align}
Thus, by Eqs.\eqref{sigma}, \eqref{connection}, \eqref{com}, \eqref{extra1}, \eqref{dphi} and \eqref{extra2}, the symplectic current on $\Delta$ reduces to
\begin{align}\label{current2}
\pback{J}(p,\delta_1,\delta_2)\ \widehat{=}\ &\frac{1}{4\pi G}\delta_{\left[1\right.}\prescript{2}{}\epsilon\wedge\delta_{\left.2\right]}(\phi w
+\frac{\i}{\gamma}V)\n\\
&+\frac{1}{4\pi G}\delta_{\left[1\right.}(\kappa_{(l)}\prescript{3}{}\epsilon)\ \delta_{\left.2\right]}\phi. 	
\end{align}
Note that only the components along $n_a$ in the expressions \eqref{com-ome} and \eqref{com-V} of $\omega_a$ and $V_a$ contribute in the right hand side of Eq.~\eqref{current2}. To describe the contribution of $V_a$, one can define a potential $\mu$ via $\mathcal{L}_l\mu\,\widehat{=}\,\i(\epsilon-\bar{\epsilon})$. Using $\mu$ and the potential $\psi$ defined in the property (c) of WIH, Eq.~\eqref{current2} can be expressed as
\begin{align}\label{current3}
  \pback{J}(p,\delta_1,\delta_2)\ \widehat{=}\ &\frac{1}{4\pi G}\delta_{\left[1\right.}\prescript{2}{}\epsilon\wedge\delta_{\left.2\right]}(\phi \d\psi+\frac{1}{\gamma}\d\mu)\n\\
  &-\frac{1}{4\pi G}\delta_{\left[1\right.}\big(\kappa_{(l)}(\d v)\wedge\prescript{2}{}\epsilon\big)\ \delta_{\left.2\right]}\phi. 	
\end{align}
where $v$ is the parameter for the integral curve of $l^a$ such that $n_a=-(dv)_a$. Taking account of the fact that $d(\prescript{2}{}\epsilon)=0$ on $\Delta$ and the condition (iii) of WIH, the symplectic current \eqref{current3} can be written as
\begin{align}\label{current4}
  \pback{J}(p,\delta_1,\delta_2)\ \widehat{=}\ \d j(p,\delta_1,\delta_2),
\end{align}
where the 2-form $j$ on $\Delta$ is given by
\begin{align}\label{surface-current}
  j(p,\delta_1,\delta_2)=&\frac{1}{4\pi G}\delta_{\left[1\right.}\prescript{2}{}\epsilon\delta_{\left.2\right]}(\phi\psi+\frac{1}{\gamma}\mu)\n\\ &-\frac{1}{4\pi G}\delta_{\left[1\right.}(\kappa_{(l)} v\,\prescript{2}{}\epsilon)\ \delta_{\left.2\right]}\phi.
\end{align}
Together with Eqs.~\eqref{current1}, \eqref{current4} and \eqref{surface-current}, Eq.~\eqref{intJ} gives the conserved symplectic form
\begin{align}
&\Omega(\delta_1,\delta_2)\n\\
=&\frac{1}{8\pi G}\int_M\!\! \delta_{\left[1\right.}\!\Sigma^{IJ}\!\!\wedge\delta_{\left.2\right]}(\phi\bar{\omega}_{IJ} \!+\!\frac{1}{\gamma}\prescript{\star}{}{\bar{\omega}}_{IJ}\!+\!e_I e^f_J\partial_f\phi) \n\\
&+\frac{1}{8\pi G}\int_M \delta_{\left[1\right.}\phi \delta_{\left.2\right]}\left(\frac{\omega(\phi)}{\phi}\prescript{\ast}{} \d\phi\right.\n\\
&-\left.\frac{1}{3}\varepsilon_{IJKL}(\nabla_d e^{dI})e^J\wedge e^K\wedge e^L\right) \n\\ &+\frac{1}{4\pi G}\int_H\delta_{\left[1\right.}\prescript{2}{}\epsilon\delta_{\left.2\right]}(\phi\psi+\frac{1}{\gamma}\mu)\n\\ &+\frac{1}{4\pi G}\int_H\delta_{\left[1\right.}\phi\delta_{\left.2\right]}(\kappa_{(l)} v \,\prescript{2}{}\epsilon).\label{sym-form}
\end{align}
Note that $\Omega(\delta_1,\delta_2)$ is actually a pre-symplectic form on the covariant phase space, since it has degenerate directions. However, it can be corresponded to the symplectic form on the phase space.
Eq.~\eqref{sym-form} shows that, for the spacetimes admitting a WIH as internal boundary, the symplectic form contains the bulk part and boundary part. The conjugate pairs for the geometry part and the scalar field part can be easily read from Eq.~\eqref{sym-form} respectively. Let $\psi=\kappa_{(l)} v$, which can always be satisfied by adjusting the initial values of $\psi$ and $v$ on $H^-$. Then except for the term containing $\gamma$, the other terms in the boundary symplectic form in Eq.~\eqref{sym-form} can be combined into that in \cite{Ashtekar:2003jh}. However, our expression indicates that the scalar field has its own degrees of freedom. Moreover a term of Immirzi parameter is introduced.

\section{The first law}
To set up the thermodynamics of WIH, one needs to define the quasilocal notions of energy, entropy, angular momentum and so on for the horizon. It turns out that these notions can be obtained naturally by asking for the existence of a consistent Hamiltonian evolution in the covariant phase space $\Gamma$ associate to $\mathcal{M}$~\cite{Lee:1990nz,Ashtekar:1991}.
Given a vector field $t^a$ satisfying proper boundary condition on $\mathcal{M}$, it can define a vector field $\delta_t$ on $\Gamma$ by $\delta_t:=(\mathcal{L}_t e,\mathcal{L}_t\bar{\omega},\mathcal{L}_t\phi)$, which satisfies the linearized equations of motion. $\delta_t$ will be a Hamiltonian vector field, if it preserves the symplectic form, \textit{i.e.}, $\mathcal{L}_{\delta_t}\Omega=0$ on $\Gamma$. The necessary and sufficient condition for this requirement is that there exists a function $H_t$  such that $\delta H_t=\Omega(\delta,\delta_t)$ for all vector field $\delta$ in $\Gamma$~\cite{Ashtekar:2000hw}. This condition is equivalent to that the one-form $X_t$ on $\Gamma$ defined by
 \begin{align}\label{eq:dH}
 X_t(\delta):=\Omega(\delta,\delta_t)
 \end{align}
  is closed, \textit{i.e.}, $\mathrm{d}\hspace{-0.97ex}\mathrm{d}X_t=0$,
where $\mathrm{d}\hspace{-0.97ex}\mathrm{d}$ denotes the exterior derivative on $\Gamma$. If this condition is satisfied, up to an additive constant, the Hamiltonian function is given by
 \begin{align}
 \mathrm{d}\hspace{-0.97ex}\mathrm{d}H_t= X_t.
 \end{align}
Suppose that the internal boundary of a spacetime region $\mathcal{M}$ is a WIH $(\Delta,[l])$ with a rotational symmetry. It is convenient to introduce a fixed rotational vector field $\varphi^a$ on $\Delta$ and admit only those spacetimes in $\Gamma$ which have this $\varphi^a$ as the horizon symmetry. Thus the geometry restricted to $\Delta$ and the scalar field $\phi$ are both Lie dragged by $\varphi^a$. Moreover, we ask $\varphi^a$ to be tangent to $H$ and have closed circular orbits of affine length $2\pi$. Consider a vector field $t^a$ on $\mathcal{M}$, which approaches an asymptotic time translation at infinity and reduces to a symmetry on $\Delta$ such that
\begin{align}
  t^a\widehat{=}B_{(l,t)}l^a-\Omega_{(t)}\varphi^a \label{eq:t}
\end{align}
for some constants $\Omega_{(t)}$ and $B_{(l,t)}$, where $B_{(l,t)}l^a$ is unchanged under the rescalings of $l\in[l]$, and $\Omega_{(t)}$ will be referred as the angular velocity of $\Delta$ related to $t^a$. Note that, while the variations $\delta_t$ of the fields $(e,\bar{\omega},\phi)$ induced by $t^a$ are Lie derivatives along $t^a$, the actions of $\delta_t$ on the parameter $v$ and the potentials $\psi$ and $\mu$ are not the same case. To preserve the initial assignment of $(\psi,\mu,v)$, the variations in $\Delta$ should satisfy $\delta_t\psi=\delta_t\mu=\delta_t v=0$ \cite{Ashtekar:2001is}.
To analyze whether $\delta_t$ is a Hamiltonian vector field on $\Gamma$, we need to use \eqref{sym-form} to calculate $X_t(\delta)=\Omega(\delta,\delta_t)$. The boundary part in \eqref{sym-form} gives
\begin{align}
&\Omega|_H(\delta,\delta_t)\n\\
=&\frac{1}{8\pi G}\int_H\delta\prescript{2}{}\epsilon(\mathcal{L}_{t}\phi)\psi -(\mathcal{L}_{t}\prescript{2}{}\epsilon)\delta(\phi\psi+\frac{1}{\gamma}\mu)\n\\
&+\frac{1}{8\pi G}\int_H\delta\phi\mathcal{L}_{t}(\kappa_{(l)} \,\prescript{2}{}\epsilon)v -(\mathcal{L}_t\phi)\delta(\kappa_{(l)} v \,\prescript{2}{}\epsilon)\n\\
=&0,\n
\end{align}
where we used the above requirement for the variations of $(\psi,\mu,v)$ on $\Delta$ in the first step, and we replaced $t$ by Eq.~\eqref{eq:t} and used the condition (iii), the properties (a), (b), and the geometric symmetry of the WIH in the last step. Therefore, the boundary symplectic form in Eq.~\eqref{sym-form} does not contribute to $X_t(\delta)$. To calculate the bulk integral, we need use the Stokes' theorem and the identity $\mathcal{L}_t\underline{u}=t\cdot d\underline{u}+d(t\cdot\underline{u})$ for a form $\underline{u}$, where $d$ is the exterior derivative on spacetime.
This identity takes the following specific forms
\begin{align}
  \mathcal{L}_{t} \bar{\omega}&=t \cdot F+D(t \cdot \bar{\omega}),\n\\
  \mathcal{L}_{t} e &=t \cdot D e+D(t\cdot e)-(t\cdot\bar{\omega}) e, \n\\
  \mathcal{L}_{t} \Sigma&=t \cdot D \Sigma+D(t \cdot \Sigma)-[(t \cdot \bar{\omega}), \Sigma],\n
\end{align} where $D$ is the internal exterior derivative defined by $\bar{\omega}$, and $[\cdot,
\cdot]$ denotes the commutator of internal indices.
Then, using the equations of motion and their linearized version,
the bulk symplectic form in Eq.~\eqref{sym-form} will contribute only two surface terms to $X_t(\delta)$, \textit{i.e.}, the variation of the ADM energy $\delta E_{\mathrm{ADM}}^t$ at spatial infinity and the variation of the horizon energy $\delta E^t_{\Delta}$ at $H$. For the ADM energy of STT, we refer to Ref.~\cite{Dyer:2008hb}. Since our aim is the first law of WIH, let us focus on the term at the horizon,
\begin{align}
  &\Omega|_H(\delta,\delta_t)\n\\
  =&\frac{1}{16\pi G}\int_H(\delta\Sigma^{IJ})(\phi t\cdot\bar{\omega}_{IJ}\!+\!\frac{1}{\gamma}t\cdot\prescript{\star}{}{\bar{\omega}}_{IJ}+t\!\cdot\! e_I e^f_J\partial_f\phi)\n\\
  &\!-\!\frac{1}{16\pi G}\int_H t\cdot\Sigma^{IJ}\wedge\delta(\phi\bar{\omega}_{IJ} +\frac{1}{\gamma}\prescript{\star}{}{\bar{\omega}}_{IJ}+e_I e^f_J\partial_f\phi)\n\\
  &+\frac{1}{16\pi G}\int_H(\delta\phi)\big(\frac{\omega(\phi)}{\phi}t\cdot\prescript{\ast}{}\d\phi-2\kappa_{(t)}(t\cdot n)\prescript{2}{}{\epsilon}\big)\n\\
  =&\frac{1}{8\pi G}\int_H\big(\phi\kappa_{(t)} +\frac{\i}{\gamma}(\epsilon-\bar{\epsilon})\big)\delta\prescript{2}{}{\epsilon} +\frac{1}{8\pi G}\int_H\kappa_{(t)}\prescript{2}{}{\epsilon}\delta\phi\n\\
  &-\frac{\Omega_t}{8\pi G}\int_{H}(\phi\varphi\!\cdot\! w
  +\frac{\i}{\gamma}\varphi\!\cdot\! V)\delta\prescript{2}{}{\epsilon}\n\\ &+\frac{\Omega_t}{8\pi G}\int_{H}\varphi\cdot\prescript{2}{}{\epsilon}\wedge\delta(\phi w+\frac{\i}{\gamma}V)\n\\
  =&\int_H\frac{\kappa_{(t)}}{8\pi G}\delta(\phi\,\prescript{2}{}\epsilon) +\Omega_t\delta J^{\varphi}_{\Delta} +\int_H\frac{\i}{8\pi G\gamma}(\epsilon-\bar{\epsilon})\delta\prescript{2}{}\epsilon,
\end{align}
where in the second equality we used the fact that the terms containing $t\cdot e_I e^f_J\partial_f\phi$ or $l\cdot\Sigma^{IJ}$ vanish by the contraction of internal indices and $\pback{\prescript{\ast}{}\d\phi}=0$, in the third equality we used the fact that the restriction of $\varphi^c w_{a}\wedge\epsilon_{bc}$ to $H$ vanishes and defined
\begin{align}\label{ang1}
J^{\varphi}_{\Delta}=-\frac{1}{8\pi G}\int_{H}(\phi\varphi\cdot w+\frac{\i}{\gamma}\varphi\cdot V)\prescript{2}{}{\epsilon}.
\end{align}
Note that Eq.~\eqref{com-V} implies $\varphi\cdot V=\varphi^a\big((\alpha-\bar{\beta})m_a+(\beta-\bar{\alpha})\bar{m}_a\big)$. Note also that we can require $m^a$ to be Lie dragged by $l$, \textit{i.e.}, $\mathcal{L}_l m^a=(\epsilon-\bar{\epsilon})m^a\widehat{=}0$. This requirement can be realized by an appropriate spin transformation to set the spin coefficient $\epsilon$ to real numbers~\cite{Ashtekar:2000hw}. Thus we obtain
\begin{align}\label{xt}
  X_t(\delta)=\frac{\kappa_{(t)}}{8\pi G}\int_H\delta(\phi\,\prescript{2}{}\epsilon) +\Omega_t\delta J^{\varphi}_{\Delta}-\delta E_{\mathrm{ADM}}^t.
\end{align}
The existence of a Hamiltonian $H_t$ is equivalent to $\mathrm{d}\hspace{-0.97ex}\mathrm{d}X_t=0$, which can be satisfied if and only if
\begin{align}\label{dkappa}
  \frac{1}{8\pi G}\mathrm{d}\hspace{-0.97ex}\mathrm{d}\kappa_{(t)}\wedge\hspace{-1.7ex}\wedge\, \mathrm{d}\hspace{-0.97ex}\mathrm{d}\left(\int_H(\phi\,\prescript{2}{}\epsilon)\right) +\mathrm{d}\hspace{-0.97ex}\mathrm{d}\Omega_t\wedge\hspace{-1.7ex}\wedge\ \mathrm{d}\hspace{-0.97ex}\mathrm{d}J^{\varphi}_{\Delta}=0.
\end{align}
The condition \eqref{dkappa} implies that $\kappa_{(t)}$ and $\Omega_t$ are only the functions of $\int_H(\phi\,\prescript{2}{}\epsilon)$  and $J^{\varphi}_{\Delta}$ respectively. As argued in Refs.~\cite{Ashtekar:2001is} and \cite{Ashtekar:2003jh}, there do exist the vector fields $t^a$ such that the condition \eqref{dkappa} is satisfied.
Then, the Hamiltonian can be written as
\begin{align}
  H_t=E_{\Delta}^t-E_{\mathrm{ADM}}^t,
\end{align}
where $E_{\Delta}^t$ is defined as the energy of WIH related to $t^a$ and given by
\begin{align}\label{1stlaw}
  \delta E^t_{\Delta}=\frac{\kappa_{(t)}\hbar c^3}{2\pi k_{\rm B}}\delta S+\Omega^t\delta J_{\Delta}^{\varphi},
\end{align}
with the entropy
\begin{align}
  S&=\frac{k_{\rm B}}{4G\hbar c^3}\oint \phi\,\prescript{2}{}\epsilon. \label{ent}
\end{align}
Here $J_{\Delta}^{\varphi}$ is defined as the angular momentum of WIH related to $\varphi^a$. In contrast to those expressions in Refs.~\cite{Ashtekar:2001is} and \cite{Ashtekar:2003jh}, where the Immirzi parameter was not included in the actions, our expression \eqref{ang1} shows that the Immirzi term does enter the definition of $J_{\Delta}^{\varphi}$. However, similar to the situation in GR\cite{Corichi:2010ur}, the total angular momentum $J_{\infty}$ at infinity will not contain the Immirzi term in STT. For the spacetime admitting a global rotational Killing vector field which reduces to $\varphi^a$ on $\Delta$, $J^{\varphi}_{\Delta}$ can match $J_{\infty}$, since the Immirzi term disappears at infinity because of the asymptotically flat condition. Eq.~\eqref{1stlaw} is our first law of WIH in STT. If one further required $[\mathcal{L}_l,\mathcal{D}]V=0$, for all vector fields $V$ tangential to $\Delta$ and all $l\in[l]$, the WIH $(\Delta,[l])$ would become isolated horizon (IH). Even the requirement for $l$ in IH is stronger than that in WIH, IH is still a non-trivial extension of the Killing horizon\cite{Chrusciel:1992rv}.  For a given non-expanding horizon $\Delta$~\cite{Ashtekar:2000hw}, in principle infinitely many WIH can be defined by it~\cite{Ashtekar:2000hw}, but generically only one IH can be defined\cite{Ashtekar:2001jb}. Since the IH satisfies stronger conditions than those of the WIH, our first law for WIH is also valid for IH, but the $l^a$ and hence $t^a$ of the latter are more strict than those of the former.

\section{Conclusion}
 To summarize, based on the first-order action \eqref{action-2} in STT, we derived the symplectic form \eqref{sym-form} for the spacetimes admitting a WIH as internal boundary via the covariant phase space approach. By asking for the existence of a consistent Hamiltonian evolution in the covariant phase space, the first law of thermodynamics for WIH with rotational symmetry was obtained as Eq.~\eqref{1stlaw}. It turns out that, besides the area of WIH, the scalar field of STT contributes also to the entropy. Our expression \eqref{ang1} for the angular momentum of WIH includes the Immirzi term in addition to the expression in Ref.~\cite{Ashtekar:2003jh}. In contrast to the action employed in Ref.~\cite{Ashtekar:2003jh}, our action \eqref{action-2} for STT is more general and includes the Immirzi parameter. Moreover, we use the Jordan frame to express the symplectic form rather than the Einstein frame. Thus, our expressions of symplectic form and angular momentum are different form those in Ref.~\cite{Ashtekar:2003jh}. The appearance of the Immirzi parameter in the expression for angular momentum of WIH is an interesting issue which deserves further studying.

 It should be noted that, based on the Wald formula \cite{PhysRevD.48.R3427} and the first-order action of GR, the first law of the thermodynamics for the Killing horizon was derived in Ref.~\cite{Jacobson:2015uqa} and then extended to general gauge invariant gravitational theories \cite{Prabhu:2015vua}. Thus, it would be interesting to compare our treatment for WIH to those treatments for Killing horizon in the first-order scalar-tensor theories. This issue is left for future study. Note also that the no-hair theorem still holds for the event horizon of a black hole in the general scalar-tensor theories~\cite{Sotiriou:2011dz,Faraoni:2017ock}. Thus, there is no scalar hair for the stationary black holes under certain conditions. However the WIH that we considered is quasi-locally defined, which does not require any global structure of spacetime. Hence our conclusion that the non-minimally coupled scalar $\phi$ would contribute to the IH entropy and angular momentum does not contradict the no-hair theorem.

Since the scalar field has its own degrees of freedom in the boundary symplectic form in Eq.~\eqref{sym-form} and appears in the entropy formula \eqref{ent} of STT, it should also contribute to the microstate number if one calculates the entropy at quantum gravity level. The results in this paper lays a classical foundation to study the statistic orgin of the black hole entropy, since the loop quantization of STT has been performed~\cite{PhysRevD.84.104045}.

\begin{acknowledgements}
The authors would like to thank Abhay Ashtekar for helpful discussion. This work is supported by NSFC with Grants No. 11875006 and No. 11961131013, the NSF grant PHY-1806356 and the Eberly Chair funds of Penn State.
\end{acknowledgements}


\end{document}